\input mtexsis
\input psfig
\def\vss{\vskip 0pt plus 1fil minus 1fil}

\def\0{\hphantom{0}}
\def\naive{na\kern0.05em\"\i ve}
\def\gl{\hbox{\raise.5ex\hbox{$>$}\kern-.8em\lower.5ex\hbox{$<$}}}
\def\gla{{\hbox{\raise.3ex\hbox{${\scriptscriptstyle >}$}%
\kern-.5em\lower.3ex\hbox{${\scriptscriptstyle <}$}}}}
\def\vol#1{{\bf #1}}
\def\MeV{\mathop{\rm Me\kern-0.1em V}\nolimits}

\def\tr{{\rm tr}}

\def\HeadLine{{\hfill}}
\def\FootLine{\ifnum1=\pageno{\hfill}\else
{\hfill{\twelverm\folio}\hfill}\fi}
\doublespaced
\superrefsfalse
\preprint
\pubdate{\twelvepoint December, 1992}
\pubcode{\twelvepoint IFUP-TH 31/92}
\titlepage
\title
\sixteenpoint\bf
Lattice Perturbation Theory by Computer Algebra:
A Three-Loop Result for the Topological Susceptibility
\endtitle
\authors
\sixteenpoint\rm
B. All\'es$^{\rm a}$, M. Campostrini$^{\rm b}$, A. Feo$^{\rm b}$,
H. Panagopoulos$^{\rm b,c}$
\fourteenpoint\rm
\institution{a}{Departamento de F\'\i sica Te\'orica y del Cosmos,
Universidad de Granada, Spain}
\institution{b}{I.N.F.N. and Dipartimento di Fisica dell'Universit\`a,
Pisa, Italy}
\institution{c}{Dept. of Natural Sciences, Univ. of Cyprus, Nicosia,
Cyprus}
\endauthors
\abstract We present a scheme for the analytic computation of
renormalization functions on the lattice, using a symbolic manipulation
computer language.  Our first nontrivial application is a new three-loop
result for the topological susceptibility.
\endabstract
\endtitlepage
\referencelist
\reference{tHooft}
G.~'t Hooft, Phys. Rev. Lett. {\bf 37} (1976) 8;
Phys. Rev. {\bf D14} (1976) 3432
\endreference
\reference{Witten}
E.~Witten, Nucl. Phys. {\bf B156} (1979) 269
\endreference
\reference{Veneziano}
G.~Veneziano, Nucl. Phys. {\bf B159} (1979) 213
\endreference
\reference{Crewther}
R.~J.~Crewther, La Rivista del Nuovo Cimento,
serie 3, Vol. 2, 8 (1979)
\endreference
\reference{DiVecchia}
P.~Di~Vecchia, K.~Fabricius, G.~C.~Rossi and G.~Veneziano,
Nucl. Phys. {\bf B192} (1981) 392;
Phys. Lett. {\bf 108B} (1982) 323
\endreference
\reference{Campostrini2}
M.~Campostrini, A.~Di~Giacomo, H.~Panagopoulos, \PL \vol{212B} (1988) 206
\endreference
\reference{Campostrini5}
M. Campostrini, A. Di Giacomo, H. Panagopoulos, E. Vicari,
Nucl. Phys. {\bf B329} (1990) 683
\endreference
\reference{Campostrini7}
M. Campostrini, A. Di Giacomo, Y. G\"und\"u\d c, M. P. Lombardo, H.
Panagopoulos, R. Tripiccione, Phys. Lett. {\bf B252} (1990) 436
\endreference
\reference{Teper3} M. Teper, Nucl. Phys. B (Proc. Suppl.)
{\bf 20} (1991) 159, and references therein
\endreference
\reference{Bitar}
K. Bitar et al., Nucl. Phys. B (Proc. Suppl.) {\bf 20} (1991) 390
\endreference
\reference{Alles}
B. All\'es, A. Di Giacomo, Phys. Lett. {\bf B294} (1992) 269
\endreference
\reference{DiGiacomo5} A.~Di~Giacomo, E.~Meggiolaro, H.~Panagopoulos,
Phys. Lett. {\bf B277} (1992) 491
\endreference
\reference{Campostrini8}
M. Campostrini, P. Rossi, Phys. Lett. {\bf 242B} (1990) 81
\endreference
\reference{Alles2}
B. All\'es, M. Giannetti, Phys. Rev. {\bf D44} (1991) 513
\endreference
\reference{Cvitanovic}
P. Cvitanovic, Phys. Rev. {\bf D14} (1976) 1536
\endreference
\reference{Veltman}
G.~'t Hooft, M.~Veltman, Nucl. Phys. {\bf B153} (1979) 365
\endreference
\reference{Passarino}
G.~Passarino, M.~Veltman, Nucl. Phys. {\bf B160} (1979) 151
\endreference
\reference{Panagopoulos}
H.~Panagopoulos, E.~Vicari, Nucl. Phys. {\bf B332} (1990) 261
\endreference
\reference{Kawai}
H.~Kawai, R.~Nakayama, K.~Seo, Nucl. Phys. {\bf B189} (1981) 40
\endreference
\reference{Luscher2}
M.~L\"uscher, P. Weisz, Nucl. Phys. {\bf B266} (1986) 309
\endreference
\reference{Novikov}
V. A. Novikov, M.~A.~Shifman, A.~I.~Vainshtein, V.~I.~Zakharov,
Nucl. Phys. {\bf B174} (1980) 378; {\bf 237} (1984) 525; {\bf 249} (1985)
445
\endreference
\reference{Alles4}
B. All\'es, M. Campostrini, A. Feo, H. Panagopoulos, {\it The Three-Loop
Lattice Free Energy}, Pisa preprint IFUP-TH 32/92
\endreference
\endreferencelist
\offparens
\section{Introduction}

The formulation of quantum field theories on the lattice has been mainly
motivated by the need to study observables which are not amenable to a
perturbative treatment. Yet, since the first days of the lattice, it became
clear that perturbation theory could not be completely done away with; many
quantities of physical interest measured on the lattice are connected to
their continuum counterparts through renormalization functions which, in
most cases, can only be calculated perturbatively. At a time when Monte
Carlo numerical results are becoming increasingly accurate, higher order
calculations of these functions, leading to non-negligible corrections, are
necessary to achieve a matching precision.

In the present paper we report on a scheme which we have developed for
doing perturbative calculations on the lattice, using a symbolic computer
language. Various schemes for doing similar calculations in the continuum
exist since many years now, starting with Veltman's Schoonschip; on the
lattice, the lack of Lorentz invariance and the non-polynomial nature of
the action introduce several additional complications, which we will point
out below. We are currently working in formulating our computational scheme
into a package for general use; in what follows we will limit ourselves to
highlighting the essential points, deferring a detailed presentation of our
algorithms to a future publication.

As a first nontrivial application we also present the calculation to three
loops of the additive renormalization (perturbative tail) of the
topological susceptibility\refrange{tHooft}{Crewther}. This operator has
been studied for a number of years by several groups, using different
methods\refrange{DiVecchia}{Teper3}, and is currently still under
investigation, in particular in actions with dynamical
fermions\refrange{Bitar}{Alles} and around finite temperature phase
transitions\ref{DiGiacomo5}; both the presence of a phase transition, and
the need to test further for agreement among the methods adopted, call for
more precise determinations of this operator.

A final introductory remark is in order here: In dealing with perturbation
theory, one must bear in mind some well-founded caveats, stemming from the
asymptotic,\nobreak ~non-Borel summable nature of the perturbative series.
As an example, the task of subtracting
\vfill\eject
\noindent
additive renormalizations (mixing with lower dimensional operators) from a
Monte Carlo signal seems rather problematic in principle.  While general
(non-)feasibility proofs are lacking, there do exist demonstrations, at
least in 2-dimensional vector models, that some of these problems can be
circumvented\ref{Campostrini8}; thus, for example, a perturbative tail can
be consistently defined and unambiguously separated from the physical
signal. At any rate, it goes without saying that consistency checks are
very important in these calculations, to ascertain that numerical results
do show the expected theoretical behaviour; fortunately, this has been the
case with most observables considered so far.

\section{Lattice Perturbation Theory by Computer}

The tasks one must carry
out in doing lattice perturbation theory on a computer (or otherwise) are,
in a nutshell:

\item{$\alpha$)} Computing the vertices

\item{$\beta$)} Generating all relevant diagrams (with correct weights)

\item{$\gamma$)} Performing the contractions for each diagram

\item{$\delta$)} Extracting powers of external momenta from the resulting
$n$-point function

\item{$\epsilon$)} Producing numerical code for loop integrations.

These tasks are independent of one another; in particular, one may choose
to perform only some of them symbolically and the rest by hand. Although
many of the issues involved are a standard part of lattice perturbation
theory, we will highlight them in a way that points out to their
algorithmic resolution. We will draw
examples from the topological susceptibility, defined as:
$$\chi=\int
d^4x\,\langle0\left|T(Q(x)Q(0))\right|0\rangle,\EQN $$
$Q(x)$ being the topological charge density:
$$ Q(x) = {g^2\over64\pi^2}\epsilon^{\mu\nu\rho\sigma}F_{\mu\nu}^a(x)
F_{\rho\sigma}^a(x).\EQN $$
Using a lattice version $Q_L(x)$ of $Q(x)$:
$$Q_L=\!\!\sum_{\mu,\nu,\rho,\sigma\gla
0}\!\!\varepsilon_{\mu\nu\rho\sigma}\,\tr(U_{x,\mu\nu}U_{x,\rho\sigma}),
\quad \left(U_{x,\mu\nu}= U_{x,x{+}\mu}\,U_{x{+}\mu,x{+}\mu{+}\nu}\,
U_{x{+}\mu{+}\nu,x{+}\nu}\,U_{x{+}\nu,x}\right)$$
$$ Q(x)= \lim_{a{\to}0}{-1\over a^4} {1\over 2^9 \pi^2} Q_L\ ,\EQN $$
$\chi$ can be obtained from Monte Carlo data of $Q_L$ through:
$${1\over2^{18}\pi^4}\sum_n\langle0
\left|Q_L(n)Q_L(0)\right|0\rangle_{\rm M.C.} =
\chi\,a^4\,\left(Z(\beta)\right)^2+\langle{g^2\over4\pi^2}F_{\mu\nu}^a
F_{\mu\nu}^a \rangle a^4\,b(\beta)+ d(\beta)\EQN fit$$
The first nontrivial loop calculations of $Z(\beta), b(\beta), d(\beta)$
were done in Refs.\refrange{DiVecchia}{Campostrini5},\ref{Alles2}. In
this work we calculate the 3-loop coefficient
$d_4$ of $d(\beta){=} d_3/\beta^3 {+} d_4/\beta^4 {+}\cdots$\ .

\medskip
Let us go briefly through the chief points in the above tasks.

\medskip

$\alpha$)\ \ Since both the action and most operators on the lattice are
written in terms of link variables, which are exponentials of the gauge
potential, they contain vertices with an arbitrarily large number of gluon
lines. The size of these vertices grows in principle as
${n{+}l{+}1\choose n}$, where $n$ is the number of gluon lines
and $l$ is the number of links in the corresponding operator; to give a
rough idea, already at sixth order the vertex for the topological charge
occupies several dozens of output pages. Thus, any algorithm for generating
vertices must take great care to keep them compact.

Generically, an $n$-point vertex can be written in the form (for the sake
of simplicity, we shall omit throughout this presentation our treatment of
ghosts and fermions; these present some further complications, but no real
stumbling blocks):
$$ {1\over (2\pi)^{4n}}\int d^4k_1\ldots d^4k_n
\sum_{\mu_1,\ldots,\mu_n \atop a_1,\ldots,a_n}
\left(\prod_{i=1}^n A_{\mu_i}^{a_i}(k_i)\,e^{-ik_i{\cdot}\mu_i/2}\right)\,
V_{\mu_1,\ldots,\mu_n}^{a_1,\ldots,a_n}(k_1,\ldots,k_n)
\EQN $$
$A_\mu^a(k)$ is the gluon field with momentum $k$ and Lorentz (color) index
$\mu\ (a)$. The phases $\exp(-ik{\cdot}\mu/2)$ ($\mu$ stands also for the
unit 4-vector in the $\mu$ direction) are absent when the fields
$A_\mu^a(k)$ are taken to reside on the center of the link; otherwise, they
are explicitly pulled out of the vertex to ensure that Hermitian operators
lead to vertices with $V$'s which are real (aside from an overall
prefactor).  Since such phases are carried by all vertices, including the
propagator, they cancel out upon contraction in any gauge; thus all
dependence on the location of each field within a link disappears.

We obtain $V$ either by a straightforward expansion of the exponentials or
by iterative use of the Baker-Campbell-Haussdorff (BCH) formula
$$ e^A\,e^B = e^{A+B+{1\over2}[A,B]+\cdots}\EQN $$
which results in terms with more definite symmetry properties.  Due care
must be taken to assign different dummy indices and momenta to multiple
powers of the fields; this is only one of many aspects of the computation
which are trivial by hand, but not in an automatic evaluation on a
computer.  We put $V$ in the form
$$ \eqalign{V = \ &C_1\left(L_1(E_1+E_2+\ldots)
+L_2(E_3+E_4+\ldots)+\ldots\right)\cr &+
C_2\left(L_3(E_5+E_6+\ldots)+L_4(E_7+E_8+\ldots)+\ldots\right)+\ldots\cr}
\EQN $$
where $C_i$ are `colour structures', $L_i$ are `Lorentz structures' and
$E_i$ are monomials in trigonometric functions of momentum components. As
an example, for the 3-point vertex of $Q_L$, $V$ takes the form
$$\EQNalign{&V_{\mu_1\mu_2\mu_3}^{a_1a_2a_3}(k_1,k_2,k_3) = -32\, i\,g^3\,
e^{ix{\cdot}(k_1+k_2+k_3)}\,
f^{a_1a_2a_3}\cdot\EQN \cr
&\quad\bigl\{\sum_{\rho_1}\varepsilon_{\mu_1\mu_2\mu_3\rho_1} \bigl[
2\cos(k_1{\cdot}\mu_1/2)\cos(k_1{\cdot}\mu_2/2)\cos(k_1{\cdot}\mu_3/2)
\cos(k_2{\cdot}\mu_3/2)\cos(k_3{\cdot}\mu_2/2)\sin(k_1{\cdot}\rho_1) \cr
&\qquad\qquad{+}4/3\cos(k_1{\cdot}\mu_1/2)
\cos(k_2{\cdot}\mu_2/2)\cos(k_3{\cdot}\mu_3/2)
\sin(k_1{\cdot}\rho_1/2)\sin(k_2{\cdot}\rho_1/2)
\sin(k_3{\cdot}\rho_1/2)\bigr]\cr
&\quad\ -\sum_{\rho_1\rho_2}\delta_{\mu_1\mu_2}
\varepsilon_{\mu_1\mu_3\rho_1\rho_2}
\bigl[\cos(k_3{\cdot}\mu_3/2)\cos(k_3{\cdot}\rho_2/2)
\sin((k_2{-}k_1){\cdot}\rho_2/2)
\sin(k_3{\cdot}\mu_1/2)\sin(k_3{\cdot}\rho_1)\bigr]\bigr\}\cr}$$

Beyond 3-point vertices more colour structures can arise, for example:
$\sum_c f^{a_1a_2c}f^{a_3a_4c}$ and $\sum_c d^{a_1a_2c}d^{a_3a_4c}$; even
though, in principle, a single structure, $\tr(T^{a_1}T^{a_2}\cdots
T^{a_n})$, would suffice for any vertex ($T^a$ is a generator of the gauge
group), it is preferable to use more symmetric structures for the sake of
compactness.  Lorentz structures proliferate on the lattice due to lack of
rotational invariance; they also require use of `internal' Lorentz indices
(denoted $\rho_i$) which are summed over.

At this stage it is crucial to exploit the fact that all vertices are
completely symmetric under interchange of external lines. We use this
symmetry to compactify the corresponding expressions for $V$ in three
steps: First, reduce colour structures to a minimum, e.g. put $\sum_c
f^{a_1a_3c}f^{a_2a_4c}$ in the form $\sum_c f^{a_1a_2c}f^{a_3a_4c}$;
second, use the residual symmetry of each color structure to reduce all
accompanying Lorentz structures to a minimum; and third, for each
color-times-Lorentz structure use its residual symmetry to reduce the
number of accompanying monomials to a minimum.

Some other aspects of the construction of vertices are, in brief: Using up
the symmetry under exchange of internal indices ($\rho_i$) for compactness
(this becomes more subtle when internal momenta are also present, as is the
case with the effective vertices of Fig.~1 (a,b), which are very convenient
constructs); defining new `tensors', such as $\delta_{\mu_1\mu_2\mu_3}$
($\equiv \delta_{\mu_1\mu_2}\delta_{\mu_2\mu_3}$, {\it not} summed over
$\mu_2$),
together with their lists of tensorial properties; establishing a standard,
`canonical' form for the trigonometric monomials in order to reduce their
number to a minimum. This last issue is rather nontrivial and still lacks a
satisfactory resolution; the point is that the most immediate candidate
prescriptions for a canonical form, such as using monomials with only one
trigonometric function for every direction or using only $k_i/2$ as
arguments of these functions, have precisely the opposite effect of what is
desired, leading to unmanageably large expressions.

Since, for any given operator, vertices need be constructed only once in
the beginning and then stored for subsequent use, considerations of speed
are rather marginal here; they become far more pressing in what follows.
Considerations of RAM usage are the main concern, since they determine the
feasibility of this step of the computation on a given computer.

\medskip

$\beta$)\ \ The algorithmic generation of diagrams, together with their
numerical weights, is the task which most resembles that of the continuum,
the only difference stemming from the plethora of lattice vertices. For
this reason, we shall not dwell on our approach, noting also that among the
five tasks on the outset this is the only one still feasible by hand (given
that calculations allowed by present computer capabilities can hardly reach
5 loops).

As for the numerical weight of any given diagram, we can readily compute it
from the formula:
$${\cal W} = w_{\rm exp}\cdot {\prod_i(n_i!)\,(\sum_ie_i)!\,
\prod_i(b_{ii}-1)!!\over
\prod_{i>j}(b_{ij}!)\,\prod_i(e_i!)\,\prod_i(b_{ii}!)}
\cdot {\prod_g(n_g!)\over\#_S} \EQN $$
Here, $w_{\rm exp}$ is the product of $(-1)^k/k!$ for each group of $k\ge
1$ identical vertices coming from the exponential of the action.  The index
$i$ runs over all vertices in the diagram; the $i$th vertex has a total of
$n_i$ legs, of which $e_i$ remain external, $b_{ii}$ (even) get contracted
among themselves, $b_{ij}$ get contracted against legs of the $j$th vertex.
Finally, $n_g$ is the number of vertices of type $g$ in the diagram and
$\#_S$ is the cardinality of that subgroup of the permutation group of all
identical vertices which leaves $b_{ij}$ invariant (acting simultaneously
on its rows and columns). In fact, $\#_S$ is the only quantity in
${\cal W}$ not given by a closed formula; however, it is a trivial matter
to generate it numerically from the `incidence matrix' $b_{ij}$.

The diagrams contributing to $\chi$ at 3 loops are shown in Fig. 2. Absent
from this list are circa 140 diagrams involving the 2-point vertex of $Q_L$
as well as the effective vertex of Fig. 1c, since both these vertices
vanish in the case of zero external momentum (provided no infrared
divergences are generated in this case, which we explicitly check).

\medskip

$\gamma)$\ \ From an algorithmic point of view, the contraction for a given
diagram entails the following:

\item{i)} For each vertex involved, look up the corresponding expression
for $V$ and symmetrize it partially, according to the incidence matrix;
thus, if two vertices are connected by $n$ lines ($b_{ij}{=}n$), only one
of the two need be symmetrized with respect to those lines. In a diagram
like that of Fig. 2k, this means a potential saving of a factor of $4!$ in
memory for intermediate expressions. Similar considerations apply for
$e_i\ge2$ and $b_{ii}\ge4$.

\item{ii)} Form the product of all partially symmetrized vertices, renaming
all indices (and momenta) as follows: Indices assigned to contracted legs
become internal ($\mu_i{\to}\rho_{i'}$, $a_j{\to}c_{j'}$,
$k_k{\to}p_{k'}$), and both internal and external indices are placed in
ascending order, so as to make sure that their names remain distinct.

\item{iii)} For each element of $b_{ij}$ ($i{>}j$) consider in pairs the
first $b_{ij}$ available powers of the gauge field from the i$^{\rm th}$
and the j$^{th}$ vertex (say, $A^{c_i}_{\rho_i}(p_i)$ and
$A^{c_j}_{\rho_j}(p_j)$) and substitute them with the propagator:
$${(2\pi)^4\over a^4}\,\delta(p_i{+}p_j)\,
\delta_{c_ic_j}\, {1\over({\hat p}_i^2)^2} \,
\left({\hat p}_i^2\, \delta_{\rho_i \rho_j} -
(1-\alpha) {\hat p}_{i,\rho_i} {\hat p}_{j,\rho_j}\right) \EQN $$
$$\bigl(\ {\hat p}_{i,\rho_i} = e^{i\,p_i\rho_i} -1,
\qquad {\hat p}_i^2 = \sum_{\rho{=}1}^4 {\hat p}_{i,\rho} {\hat
p}_{i,\rho}^*\ \bigr) $$
Similarly for self-contractions ($b_{ii}\ge 2$).

\item{iv)} Simplify color structures. Using the identity:
$$T^a_{ij}T^a_{kl}={1\over2}\delta_{il}\delta_{kj} -
{1\over2N}\delta_{ij} \delta_{kl}\EQN $$
(valid for $SU(N)$), all internal color indices are completely eliminated.
The algorithms which implement this simplification are identical to the
ones used in the continuum\ref{Cvitanovic}.

\item{v)} Eliminate all Kronecker $\delta$'s involving internal Lorentz
indices. Doing so requires a judicious partial expansion of the expression
(which is a product of large sums), to avoid drastic increases in memory.

\item{vi)} Compactify the expression using the symmetry under permutations
of the names of internal momenta and Lorentz indices. Allowed permutations
for momenta are those consistent with the topology of the corresponding
diagram, that is, permutations which leave invariant the momentum
conservation delta functions $\delta(p_1{+}\cdots{+}p_n)$ (or
$\exp\,ix{\cdot}(p_1{+}\cdots{+}p_n)$) at each vertex; a table of these
permutations is constructed right from the beginning. A conceptually easy
algorithm would now generate all permuted versions of each subexpression
and then select the first version in some order (e.g.  lexicographic). The
problem is that both intermediate memory and execution time will grow
factorially with the number of indices; since we often encounter up to ten
indices, already at three loops, it is clear that such an algorithm will
not do. At the price of a rather complicated source code, we have come up
with an ordering algorithm which is (quasi-)polynomial in nature.

\item{} Other considerations made here are ordering momenta simultaneously
with indices, and casting our expressions in a form involving at most four
Lorentz indices (given that the theory is defined in four dimensions).

\item{vii)} Finally, trigonometric simplifications are systematically
performed throughout, in order to put all terms in a canonical form, which
then makes identifications or cancellations automatic.

\bigskip
$\delta)$\ \ Extracting the analytic, exact momentum dependence of
an $n$-point function, in the limit $a{\to}0$, is one of the most
complicated tasks, both conceptually and algorithmically.  This task
does not enter the calculation which we present here; its elaboration is
still in progress, and will be essential for the calculation of
multiplicative renormalizations.

Even in continuum regularizations this problem is only completely resolved
at one loop\ref{Veltman},\ref{Passarino}. To arbitrary loops, no systematic
analysis of $n$-point functions exists.  On the lattice, the first step is
to decompose a given expression (to be integrated over internal momenta) in
terms of a limited set of potentially divergent integrands, plus other
terms which can be evaluated by setting the external momenta directly to
zero. A possible basis for this set is:
$${\prod_{i{=}1}^n \prod_{\mu{=}1}^4 (\sin p_{i\mu})^{\alpha_{i\mu}} \over
\prod_{i{=}1}^n({\hat p}_i^2)},\qquad
0\le\alpha_{i\mu}\le\alpha_{i\mu}^{\rm max}\EQN basis$$
This decomposition poses no conceptual difficulty, but can cause
disproportionate increases in memory, unless it is carefully implemented.
One must now integrate the above set, expressing the result in terms of
standard functions (logarithms, Spence functions, etc.) and numerical
constants characteristic of the lattice. At one loop this has been done
systematically\ref{Panagopoulos}, using a dimensional regularization
technique\ref{Kawai}. At higher loops, not only do these integrals become
quite complicated, but their number also grows significantly. We are
presently developing algorithms for carrying out the integration of the
basis functions \Ep{basis} symbolically.

\medskip
$\epsilon)$\ \ Having arrived this far, the only remaining task is the
numerical evaluation of loop integrals with no dependence on external
momenta. We do this both for finite and infinite lattices.

On a finite lattice, loop integrals become nested multiple sums (with due
attention paid to propagator zero modes). A mere conversion of the
integrand to Fortran or C syntax is almost immediate. However, to produce
optimized code one must take into account the following factors:

\item{i)} Under certain changes of variables
($\{p_{i\mu}{\leftrightarrow}p_{i\nu},\ \forall i\}$,
$\{p_{i\mu}{\to}{-}p_{i\mu},\ \forall i\}$, etc.) the integrand stays
invariant (or can be rendered invariant at a small expense in size). It
thus suffices to integrate over a small hypertriangular region of the
original domain $\{{-}\pi\le p_{i\mu}\le\pi,\ \forall i,\mu\}$. An added
complication for finite lattices is that the boundaries of this region are
sets of nonzero measure.

\item{ii)} Most diagrams contain two or more loops with no propagator line
in common. (Among three loop diagrams the `Mercedes', Fig. 2f, is the only
exception.) Integration over the corresponding loop momenta need not be
nested, but can be done independently, since all denominators
(propagators) can be factorized into terms containing at most one such
momentum. In order to factorize numerators as well, one must expand
trigonometric functions containing more than one of the above loop momenta
(again, our algorithms are written with an eye on keeping expansions to a
minimum). The computational load for the resulting code is comparable
to that of lower loop integrals.

\item{iii)} The trigonometric functions comprising each monomial in the
integrand typically depend only on a very small subset of the integration
variables and can thus be pulled out of most nested integrals. Further,
since such factors are shared among many monomials, one can organize them
in an (inverse!) tree, to avoid redundant integrations.

\smallskip
We have incorporated all the above considerations in an algorithm which
takes an integrand as input and produces optimized Fortran code for its
integration. For lattice sizes of interest (${\sim}16^4$), this
optimization results in a gain in execution time of a factor of $10^7\,$!

On infinite lattices, a drastic optimization is achieved by putting all
propagators in the Schwinger representation:
$${1\over\hat p_i^2}=\int_0^\infty e^{-\alpha_i\hat p_i^2} d\alpha_i\EQN $$
In this representation, integrations over different spatial directions
factorize, so that their effective number is reduced by $3L{-}P$ ($L$: \#
of loops, $P$: \# of propagators). At least one of the remaining
integrations can be done analytically in terms of Bessel functions, leaving
fewer integrals to be done numerically and a less singular integrand. We
illustrate this with a very simple example:
$$I\equiv \int_{-\pi}^{+\pi}{1\over\hat p^2\hat q^2\widehat{p{+}q}^2}
{d^4p\over(2\pi)^4} {d^4q\over(2\pi)^4} =\int_0^\infty d\alpha_1 \,
d\alpha_2 \, d\alpha_3 \, e^{-8(\alpha_1{+}\alpha_2{+}\alpha_3)}\,
\Phi^4(\alpha_1,\alpha_2,\alpha_3)\EQN $$
$$\eqalign{\Phi(\alpha_1,\alpha_2,\alpha_3)&=\int_{-\pi}^{+\pi}
{dp\,dq\over(2\pi)^2} \,
\exp(2\alpha_1\cos p + 2\alpha_2\cos q +2\alpha_3\cos(p{+}q))\cr &=
\int_{-\pi}^{+\pi} {dp\over2\pi} e^{2\alpha_2\cos p}\,
I_0(2\sqrt{\alpha_1^2{+} \alpha_3^2{+} 2\alpha_1\alpha_3\cos p})\cr}\EQN $$
($I_0$ is the modified Bessel function.) The resulting expressions are
amenable to high-precision Gauss-Legendre type integration.

 We also compare results for the infinite lattice to an extrapolation of
finite lattice ($L^4$) results, of the form:
$${\rm result}(L)= A+{B\over L^n}\EQN $$
In addition to $A$ and $B$, the exponent $n$ may also vary for different
diagrams (and one may also expect logarithms of $L$, cf. \Ref{Luscher2}).
The discrepancy between this extrapolation and the infinite lattice result
is typically a fraction of one per mille.

In Table I we present the results for each diagram contributing to $d_4$,
for lattices of different size. Adding up all the contributions, we obtain
for $d_4$ (on an infinite lattice):
$$d_4=N^4\,(N^2-1)\,(1.735N^2-10.82+73.83/N^2)\, 10^{-7}\EQN $$

Our programs are implemented in the computer language Mathematica. All
numerical computations are performed by separate Fortran programs generated
by our Mathematica routines. The generation of all vertices of $Q_L$ with
up to six legs (needed for the two-loop calculation of $Z(\beta)$) requires
approximately 30 hours on a SUN Sparc Station 2 with 32 Mbytes of RAM. The
computation has to be split into $\sim200$ independently computed
contributions (summed only at the very end) in order to fit intermediate
results into available RAM.

\section{Results and Conclusions}

The value obtained for $d_4$ allows one in principle to extract $\chi$ with
greater precision. With this value and those previously obtained for $d_3$
we have performed a series of best fits of \Eq{fit} to Monte Carlo data for
$SU(2)$ and $SU(3)$. As in Refs.\ref{Campostrini5},\ref{Campostrini7} we
have neglected the mixing with $\langle FF\rangle$. The values obtained for
$\chi_u$ (the non-renormalized topological susceptibility,
$\chi_u{=}\chi\,Z^2(\beta)$) are shown in Table II. They are in good
agreement with those of Refs.\ref{Campostrini5},\ref{Campostrini7}\ (q.v.\
for details). At this stage, increased statistics of Monte Carlo data would
be quite welcome on three counts: Improving the estimate of $\chi$,
assessing the importance of the $\langle FF\rangle$ mixing, and checking
for nonperturbative contributions in $d_n$ (such contributions are possible
only for sufficiently high $n$, cf. \Ref{Novikov}).

In conclusion, the calculational scheme which we have developed allows us
to perform lattice perturbative calculations automatically, with very
little `human intervention'. Our aim is to be able to repeat the
computation for different lattice operators without further programming,
and this has not yet been completely achieved.
The greatest difficulty that had to be
overcome was the existing constraints on computer time and memory, which
necessitate devising polynomial-type algorithms and optimizing every aspect
of this scheme. One major task still left open is the algorithmic
extraction of external momentum dependence. Our first original application
was the evaluation to 3 loops of the perturbative tail of $\chi$. We have
also obtained three loop results for the gluonic condensate, and report
them in a forthcoming publication\ref{Alles4}. Repeating these
calculations in the presence of dynamical fermions is relatively
straightforward: Only a few additional diagrams appear, requiring no
further computational resources.
The calculation of multiplicative
renormalizations within this scheme, as well as a technical description
of our algorithms, are postponed to a future publication.

\bigskip
\noindent
{\bf Acknowledgements.}\ \ It is a pleasure to thank Adriano Di Giacomo for
many useful conversations. We acknowledge financial support from MURST
(Italian Ministry of the University and of Scientific and Technological
Research) and from the Spanish-Italian ``Integrated Action'' (contract
A17). B.A. also acknowledges a Spanish CICYT contract.

\vfill\eject
\centerline{\bf References}
\bigskip
\doublespaced
\References

\vfill\eject
\centerline{\bf Table I}
\smallskip
\doublespaced

We list here the contribution to $d_4$ of individual diagrams, shown in
Fig. 2, in the Feynman gauge. We use an $L^4$ lattice and gauge group
SU(N). Each entry must be multiplied by $N^6(N^2-1)\,{\times}10^{-7}$.
\bigskip



\ruledtable
Fig. | $L=3$ | $L=8$ | $L=16$ | $L=\infty$  \cr
a| $31.23{-}33.75/N^2$ | $37.79{-}41.25/N^2$ | $37.79{-}41.38/N^2$ | 
   $37.77{-}41.40/N^2$  \cr
b| ${-}2.137$ | ${-}2.981$ | ${-}3.081$ | ${-}3.114$  \cr
c+d| $5.210$ | $10.07$ | $10.78$ | $11.02$  \cr
e| ${-}4.328$ | ${-}5.964$ | ${-}6.162$ | ${-}6.228$  \cr
f| $.3955$ | $.7233$ | $.7429$ | $.7457$  \cr
g| $.0882$ | $.3107$ | $.3100$ | $.3099$  \cr
h| $1.988$ | $1.063$ | $.8017$ | $.7082$  \cr
i+j| $0$ | $9.242\,{\times}10^{-6}$ | $9.112\,{\times}10^{-6}$ |
   $9.109\,{\times}10^{-6}$  \cr
k| $12.36{-}18.85/N^2$ | $17.06{-}24.49/N^2$ | $17.56{-}24.60/N^2$ |
   $17.73{-}24.61/N^2$  \crnorule
 | ${+}56.56/N^4$ | ${+}73.48/N^4$ | ${+}73.81/N^4$ | 
   ${+}73.83/N^4$ \cr
l| ${-}44.09{+}44.99/N^2$ | ${-}56.37{+}55.00/N^2$ |
   ${-}57.02{+}55.17/N^2$ | ${-}57.20{+}55.18/N^2$
\endruledtable

\bigskip\bigskip
\centerline{\bf Table II}
\smallskip

We list the values of $\chi_u/\Lambda_{\rm QCD}^4$ (for gauge groups
$SU(2)$ and $SU(3)$, as obtained from \Eq{fit} and Monte Carlo
data\ref{Campostrini5},\ref{Campostrini7} through a series of fits, in
which: a) $d_4$ was an additional parameter to be fitted, or: b) the exact
value of $d_4$ was taken from our calculation, fitting instead $d_5$.

\bigskip



\ruledtable
$\chi_u/\Lambda_{\rm QCD}^4$ | $SU(2)$ | $SU(3)$  \cr
a | $2.35(22)\,10^4$ | $2.58(64)\,10^5$  \cr
b | $2.07(23)\,10^4$ | $2.70(69)\,10^5$ 
\endruledtable

\vfill\eject

\noindent\hfill
\psfig{figure=fig1.eps,width=14truecm}
\hfill

\vskip 1truecm
\noindent\hfill
\psfig{figure=fig2.eps,width=14truecm}
\hfill
\vfill\eject
\bye